# Interaction between a moving electron and magnetic flux in Aharonov-Bohm effect

Wang    Rui-Feng

Department of Physics, Beijing Jiaotong University, Beijing 100044, China

E-mail: aboud56789@163.com

## Abstract

The back-action exerted by the moving electron on the magnetic flux in the A-B effect is analyzed. It is emphasized that a reasonable interpretation on the A-B effect should be consistent with the uncertain principle. If the back-action on the magnetic flux is reduced to zero, the A-B effect should not be observed, even through the vector potential still exists in space. To verify this interpretation, a new experimental scheme is proposed in this paper.

# Ⅰ  Introduction

The measurement in quantum mechanics is quite different from that in classical physics. In classical physics, the disturbance on the measured object can be reduced to arbitrarily small and the measurement precision has no limitation. One can measure the position and momentum of an object to arbitrary precision simultaneously. But in quantum mechanics, it is impossible to measure an object without disturbance on it, which is guaranteed by Heisenberg's uncertainty principle. The more exact the measurement, the stronger the disturbance exerted on the measured object[1,2]. If The Aharonov-Bohm effect is considered from the viewpoint of measurements, some new viewpoints must be introduced.

In 1959, Y. Aharonov and D. Bohm[3]  predicted that a phase difference between two electron beams is proportional to the magnetic flux enclosed by their paths, even though they always move in the region where the magnetic field is zero. The existence of this effect has been supported by some experiments [4], especially the experiments by A. Tonomura in 1986[5]. So the existence of the A-B effect should not be questioned. This effect has become a standard part of the quantum mechanics textbook[6]. But, the interpretations about the A-B effect can be classified into two kinds roughly: The first interpretation asserted that this effect is due to the vector potential describing the magnetic flux[7]. This interpretation is named as "the interpretation of vector potential" in this paper. The second interpretation asserted that this effect is due to the interaction energy between the moving electron's magnetic field and the magnetic flux[8,9,10]. This interpretation is named as "the interpretation of interaction energy".  The interpretation of vector potential has become a current interpretation , which states that the Aharonov-Bohm effect represents a new quantum topological effect, and the vector potential **A** can result in some observable phenomena in quantum mechanics, though it is only a mathematical field in classical physics[11]. However, the former studies on this effect always focused on the action exerted by the magnetic flux on the moving electron, but neglected the back-action exerted by the moving electron on the magnetic flux ( or on the solenoid which produces this magnetic flux). If this back-action is considered, it can be found that the interpretation of vector potential is contradictory to Heisenberg's uncertainty principle.

## Ⅱ  Back-action on the magnetic flux

The phase difference $\Delta\varphi$ between two electron beams is proportional to the magnetic flux $\Phi$ enclosed by their paths, i.e.

$$\Delta\varphi = \frac{2\pi\Phi}{h/e} \qquad (1)$$

So, by the phase difference between these two electron beams, we can measure the magnetic flux $\Phi$ produced by a solenoid. The problem is what is the disturbance on the magnetic flux $\Phi$ (or on the solenoid producing the magnetic flux $\Phi$) during this measurement process?  Or in other words, what is the back-action exerted by the moving electron beams on the magnetic flux ?

First of all, this disturbance on the magnetic flux must exist, for it is indispensible to explain the wave-particle duality, uncertainty principle and wave-packet reduction according to Copenhagen interpretation[12,13,14]. The second, this disturbance should be electromagnetic interaction. So, a natural answer to this question is the magnetic field produced by the moving electron beams will interact with the magnetic field produced by the solenoid and have a influence on the measured magnetic flux (or on the solenoid).

In more detail, as shown in Fig 1, suppose a long, straight solenoid produce a static field $\mathbf{B}_2(\mathbf{r})$ inside it and a moving electron produces a magnetic field $\mathbf{B}_1(\mathbf{r})$ around it. When the electron approaches to the solenoid, the magnetic field $\mathbf{B}_1(\mathbf{r})$ in the solenoid will increase; But when the electron leave away from the solenoid, the magnetic field $\mathbf{B}_1(\mathbf{r})$ in the solenoid will decrease. The change of the magnetic field $\mathbf{B}_1(\mathbf{r})$ will induce a electromotive force $\Delta U$ across the solenoid ( This is the back-action exerted by the moving electron beams on the magnetic flux). This analysis is consistent with the conclusions of quantum circuit, where the magnetic flux $\Phi$ through the solenoid and the voltage $U$ across the solenoid are non-commuting, *i.e.* an measurements on the magnetic flux $\Phi$ through the solenoid must be accompanied by an inevitable perturbation on its voltage $U$. (see the appendix of this paper). The mutually interaction between the moving electron and the long solenoid can be described by the interaction energy $W'$ between the magnetic field $\mathbf{B}_1(\mathbf{r})$ and $\mathbf{B}_2(\mathbf{r})$, which is given by [10]

$$W' = \int \frac{1}{\mu_0} \mathbf{B}_1 \cdot \mathbf{B}_2 d^3 r = \mathbf{A}_2(\mathbf{x}) \cdot q\mathbf{v} . \qquad (2)$$

Where, $q, \mathbf{v}$ and $\mathbf{x}$ are the charge, velocity and position of the moving electron respectively, $\mathbf{A}_2(\mathbf{x})$ is the vector potential describing the magnetic field $\mathbf{B}_2(\mathbf{r})$ inside the solenoid, which is given by:

$$\mathbf{A}_2(\mathbf{x}) = \frac{1}{4\pi} \int_\Omega \frac{\mathbf{B}_2 \times (\mathbf{x} - \mathbf{r})}{|\mathbf{x} - \mathbf{r}|^3} dr^3 \qquad (3)$$

In the literature[10], using the variational method of quantum mechanics, we have showed that the A-B effect is due to the interaction energy $W'$ between the magnetic fields $\mathbf{B}_1(\mathbf{r})$ and $\mathbf{B}_2(\mathbf{r})$, but not the vector potential $\mathbf{A}_2$ [15].

In this paper we will show that the interpretation of interaction energy on the A-B effect is consistent with the measurement theory of quantum mechanics, but the interpretation of vector potential is contradictory with it.

# Ⅲ  Disadvantages with the interpretation of vector potential

Let's consider an experiment as depicted in Fig. 2. There is a long, straight solenoid in space, which is coated by a long, straight superconducting cylinder with its thickness $d$ much larger than its penetration depth $\lambda$. A coherent beam of electrons is split into two parts, each going on opposite sides of the superconducting cylinder, but avoiding it. Then, these two beams are brought together to interfere with each other. In this experiment, the speed of the electron is less than $10^5 m/s$ and the superconducting cylinder can completely shield the electron's magnetic field outside it, due to the Messiner effect[18]. This experiment is similar to the experiment by A. Tonomura. The first difference between them is that the long, straight solenoid coated by a superconducting cylinder in this experiment replaces the tiny toroidal magnet covered entirely with a superconducting layer in the experiment by A. Tonomura. The second difference is that the speed of the electron in this experiment is less than that in the experiment by A. Tonomura.

When the superconductor cylinder is cooled into the superconducting state, the magnetic flux enclosed by it should be quantized in units of $\Phi_0 = h/2e$ [19,20]. So, for simplicity, in the beginning of the experiment, we assume that the magnetic flux $\Phi_2$ produced by the solenoid equals to $n\Phi_0$ at the temperature above the $T_C$ of the superconductor cylinder. In this case, after the superconductor cylinder is cooled into the superconducting state, no current is formed on its inner surface, so the vector potential $\mathbf{A}_2$ describing the field $\mathbf{B}_2(\mathbf{r})$ inside the solenoid still remains unchanged in the space. For a single-turn closed loop outside the superconducting cylinder, $\oint \mathbf{A}_2 \cdot d\mathbf{l} = \Phi_2 = n\Phi_0$. If the interpretation of vector potential on the A-B effect is right, which asserts that the A-B effect arises from the vector potential, then, the A-B effect should still exist in this situation. According to the equation (1), if $\Phi_2 = (2n+1)\Phi_0$, then the relative phase shift $\Delta\varphi$ between two electron beams should be : $\Delta\varphi = (2n+1)\pi$ ; But if $\Phi_2 = 2n\Phi_0$, then $\Delta\varphi = 2n\pi$. That is to say, the interference pattern of the electron beams with $\Phi_2 = (2n+1)\Phi_0$ is different from that with $\Phi_2 = 2n\Phi_0$. So, one can judge the magnetic flux enclosed by the superconducting cylinder is odd times or even times of $\Phi_0$ by the interference pattern. Obviously, this is a measurement on the magnetic flux enclosed by the superconducting cylinder. But in this process, the moving electrons cannot exert any influence on the solenoid, for the magnetic field and electronic field produced by the moving electrons have been completely shielded out of the superconducting cylinder. In Quantum mechanics, a measurement without back-action on the measured object is impossible, for it is contradictory to the uncertain principle. So, the interpretation of vector potential about this effect should be modified: the phase of a moving electron should not depend on the vector potential. Some people perhaps argue that the moving electron maybe exert an "A-B effect-type" interaction on the solenoid. A simple analysis shows that it is impossible. Suppose the magnetic field $\mathbf{B}_1(\mathbf{r})$ produced by a moving electron is described by the vector potential $\mathbf{A}_1(\mathbf{r})$, i.e. $\nabla \times \mathbf{A}_1(\mathbf{r}) = \mathbf{B}_1(\mathbf{r})$. Since $\mathbf{B}_1(\mathbf{r}) = 0$ in the region enclosed by the superconducting cylinder, where the solenoid is located, for any closed loop in this region $\oint \mathbf{A}_1 \cdot d\mathbf{l} = \int \mathbf{B}_1 \cdot d\mathbf{s} = \mathbf{0}$. Therefore, the moving electron cannot exert the "A-B effect-type" back-action on the solenoid.

So, the interpretation of vector potential on the A-B effect is not right.

# Ⅳ  Interpretation of interaction energy

We will find that the contradiction above does not exist in the interpretation of interaction energy, which attributes the A-B effect to the interaction energy between the magnetic fields, but not the vector potential **A**: After the superconductor cylinder is cooled into the superconducting state, due to the Messiner effect it can completely shield the magnetic field produced by the moving electron out of the cylinder. Consequently, the magnetic field produced by the moving electron cannot superpose with the static magnetic field in the solenoid, and the interaction energy $W'$ between them becomes zero. So, the A-B effect should not be observed any more, i.e. no matter $\Phi_2 = (2n+1)\Phi_0$ or $\Phi_2 = 2n\Phi_0$, the interference pattern between the electron beams should be same. Therefore, an observer has no way to measure the magnetic flux through the superconducting cylinder by the interference pattern of the electron beams moving outside the cylinder. Obviously, this conclusion is consistent with the measurement theory of quantum mechanics.

The further reason for this problem should be understood as following: The concept of "energy" plays a fundamental role in quantum mechanics replacing the concept of "force" in classical physics. Consequently, the electromagnetic potentials $\varphi$ and **A**, which describe the interaction energy between a charge $q$ and electromagnetic fields, appear in the Hamiltonian of the Schrödinger's equation, instead of the electric field **E** and magnetic field **B**, which describe the force acting on the charge in a electromagnetic field. Though the Aharonov-Bohm effect is connected to the electromagnetic potential in the differential equation, it does not arise from the electromagnetic potential, but from the interaction energy described by the electromagnetic potential. Only in this way to explain the A-B effect, can the conclusions be consistent with Heisenberg's uncertain principle.

## Ⅴ  Analysis on the experiment by A. Tonomura

It is necessary to point out that the superconducting layer can only confine the magnetic flux within it, but cannot shield the magnetic field produced by the electron beams in the experiment by A. Tonomura. The main reason is that the velocity of the electron beams is too fast (about $2\times10^8 m/s$ [21]) and the magnetic field $\mathbf{B}_1(\mathbf{r})$ produced by the moving electron forms a very short pulse at the tiny magnet . If the Fourier transform is introduced, the main frequency of this pulse is about $\nu \approx 5\times10^{13} Hz$, i.e. $h\nu \approx 2\times10^{-2} eV$ which is much larger than the energy gap of the Nb film (about $3\times10^{-3} eV$). Obviously, the Nb film cannot shield the magnetic field variation with so high frequency[22]. Therefore, the magnetic field $\mathbf{B}_1(\mathbf{r})$ produced by the moving electron can still penetrate the superconducting film and superpose with the static magnetic field in the tiny magnet, so, the interaction energy does not equal to zero.  Just for this reason, the A-B effect was observed in their experiments ( i.e. The interference patter with $\Phi = 2n\Phi_0$ is different from that with $\Phi = (2n+1)\Phi_0$). The detail analysis about this problem can be found in the literature[10]. If the speed of the electron beam is lowered and the superconducting layer can shield the magnetic field produced by the moving electron beams, our prediction will be supported by the experimental results.

The analysis above is consistent with the results of the preliminary experiment by A. Tonomura, which was to test whether or not there would be any observable interaction of an electron beam with a toroidal superconductor containing no magnet [ §5.5.5 of the second part in the ref [7] ]. For an incident electron passing through a field-free region has no reason to receive any force.

However, when a superconductor is located near the electron, the electron might somehow be influenced because the magnetic field produced by the passing electron cannot penetrate into the superconductor, due to the Meissner effect. The influence should result in the change of the interference patter of the electron beams. But this experimental results showed that no relative phase shift was produced between two electron beams passing both inside and outside a superconducting toroid containing no magnet. That is to say, no experiment demonstrated that the superconducting toroid can shield the magnetic field produced by the electron beams, so, our analysis above is reasonable.

## Ⅵ  A new experimental scheme

For the disadvantages with the experiment by A. Tonomura, a new experimental scheme using SQUID (Superconducting Quantum Interference Device) was proposed in the literature[10]. Here, we will improve this experimental scheme to make it more easily.

A dc-SQUID is a superconducting loop with two Josephson junctions [23], The critical current $I_C$ passing through the SQUID depends on the flux $\Phi$ through the superconducting loop, even though the magnetic field is zero in the region where the superconducting loop is located.

$$I_C = I_0 \left| \cos\left(\frac{\pi \Phi}{\Phi_0}\right) \right| \qquad (4)$$

Obviously, the SQUID and the A-B effect involve a same subject, which is the influence exerted by a magnetic field on the phase of the moving electrons. So, the physical essence of the SQUID is similar to that of the A-B effect. The only difference between them is: the moving electrons in the A-B effect are single electrons, but the moving electrons in the SQUID are Cooper pairs[24].

For a SQUID, two questions should be answered. The first question is:

what is the back-action (or the disturbance) exerted by the SQUID on the magnetic flux $\Phi$ ( or on the solenoid which creates the magnetic flux $\Phi$)? The only answer to this question is that the magnetic field produced by the current in the SQUID will exert a magnetic-action on the magnetic flux $\Phi$ ( or on the solenoid). The second question is: does the critical current $I_C$ depend on the magnetic flux $\Phi$ through the SQUID (which is equivalent to on the vector potential $\mathbf{A}$ describing the magnetic flux $\Phi$) or on the interaction energy between the magnetic flux $\Phi$ and the magnetic field produced by the SQUID? This question can be answered with a experiment as following:

The experimental arrangement is depicted in Fig 3. The dc-SQUID used in this experiment is a "point contact" device with the diameter of its superconducting loop being several centimeters  (This device was described in the literature[25], especially the Fig 7 of this reference).  Two similar solenoid denoted as *a* and *b* are enclosed by the superconducting loop of the dc-SQUID. The solenoid *a* is connected to the flux-locked feedback loop of this system and used as a negative feedback coil[26]. The function of the flux-locked feedback loop is to keep the current $I_C$ of the SQUID constant by changing the current $I_a$ in the solenoid *a*.  The solenoid *b* is coated by a superconductor cylinder outside it and connected to a independent constant-current source. The working temperature range of the dc-SQUID is from $T_2$ to $T_1$ ($T_2 < T_1$). The critical temperature $T_C$ of the superconducting cylinder lies between $T_2$ and $T_1$. The experimental steps can be performed as following:

Step 1. Let the dc-SQUID system working at the temperature $T'$ ($T_C < T' \leq T_1$). The superconductor cylinder is in normal state. The flux-locked feedback loop is turned off and the current $I_a$ in the solenoid *a* remains zero during this step. To increase the current $I_b$ in the solenoid *b* to a value ensuring that the flux through each turn of the solenoid $b$ $\Phi_b = n\Phi_0$ ( $n \gg 1$, $n$ is a integer, this request is just for simplicity in discussion).

Step 2. The flux-locked feedback loop is turned on, then to decrease the temperature of the system to $T''(T'' < T_C)$. For the temperature gradient inevitably exists in the system during the temperature decreasing, the superconducting cylinder will be cooled into the superconducting state from one end to the other step by step. In this process, the current $I_b$ in the solenoid $b$ remains unchanged, i.e. $\Phi_b = n\Phi_0$ does not change, and the vector potential $\mathbf{A}_b$ describing the magnetic field in the solenoid $b$ does not change either. But the interaction energy $W'$ between the magnetic field produced by the SQUID and the magnetic field in the solenoid $b$ becomes zero step by step.

If the critical current $I_C$ of the SQUID is dependent on the magnetic flux through its superconducting loop ( i.e. the vector potential), then the critical current $I_C$ should remain invariable in this process, for $\Phi_b$ and $\mathbf{A}_b$ remains unchanged. So, it is unnecessary to change the the current $I_a$ in solenoid $a$ to keep the $I_C$ invariable. Therefore, the magnetic flux $\Phi_a$ through each turn of the solenoid $a$ remains zero in this process. The total magnetic flux through the SQUID is: $\Phi = \Phi_a + \Phi_b = n\Phi_0$ after the whole superconductor cylinder is cooled into superconducting state.

But, If the critical current $I_C$ is dependent on the interaction energy between the magnetic field produced by the SQUID and the magnetic field through its superconducting loop, then, while the superconducting cylinder is cooled into superconducting state, the current $I_a$ in solenoid $a$ should increase to compensate for the change of the interaction energy between $I_C$ and $\Phi_b$. After the whole superconducting cylinder is cooled into superconducting state, the magnetic flux $\Phi_a$ through the solenoid *a* should equal to the $\Phi_b$ in the solenoid *b* i.e. $\Phi_a = n\Phi_0$. Only in this way can the interaction energy remain unchanged. The total magnetic flux through the SQUID is $\Phi = \Phi_a + \Phi_b = 2n\Phi_0$ after the superconducting cylinder is cooled into superconducting state. We believe that this result will appear in the experiments, showing that the A-B effect is due to the interaction energy between magnetic fields.

So, the experiment above can distinguish what is the key factor affecting the phase of the moving electrons, the electromagnetic potentials themselves or the interaction energy described by the electromagnetic potentials? The answer to this problem can help us to properly understand the Schrödinger's equation in electromagnetic fields.

# Ⅶ Conclusions

The back-action exerted by the moving electron on the magnetic flux in the A-B effect has been analyzed. It is found that the interpretation of interaction energy on the A-B effect is consistent with the uncertain principle, but the interpretation of vector potential is contradictory with it. To verify this viewpoint, a new experimental scheme is proposed in this paper.

## Acknowledgement
The author thanks Mrs. Deng Chang-yu for her encouragement.    This work is supported by "the Fundamental Research Funds for the Central Universities, No: 2009JBM102"

## Appendix A

In quantum circuit[27], for a lossless LC quantum circuit, which is composed of an inductance $L$ and a capacitance $C$, the charge $q$ on the capacitance and the variable $p$ satisfy the commutation relation $[q, p] = i\hbar$, where $p$ is given by $p(t) = L\frac{dq}{dt}$. Because the magnetic flux through the inductance $\Phi(t) = L\frac{dq}{dt} = p$, and the voltage across the inductance (or the capacitance) $U = \frac{q}{C}$, the commutation relation between $U$ and $\Phi$ is: $C[U, \Phi] = i\hbar$. ( Here, $q$, $p$, $U$, $\Phi$ are operators in quantum mechanics, $C$ and $L$ are constants ). It means that any measurement on the magnetic flux $\Phi$ through a solenoid must be with a perturbation on its voltage $U$.

[15] Based upon the equation (2), some authors [16,17] postulated the moving electron experience a non-Lorentzian force $F' = -\nabla W'$. This postulation is unnecessary and questionable. For such a force was not be observed in any classical experiment. Furthermore, As a purely quantum-mechanical phenomenon, the A-B effect should not be explained by the force concept, which has faded away in quantum physics.

FIGURES:

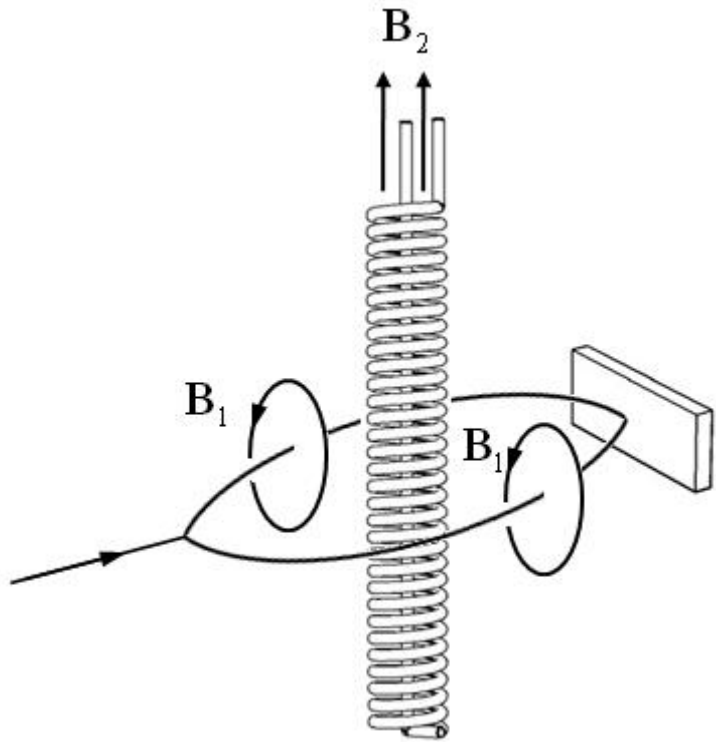

Fig. 1, The magnetic interaction between the moving electron and the magnetic flux in the A-B effect.

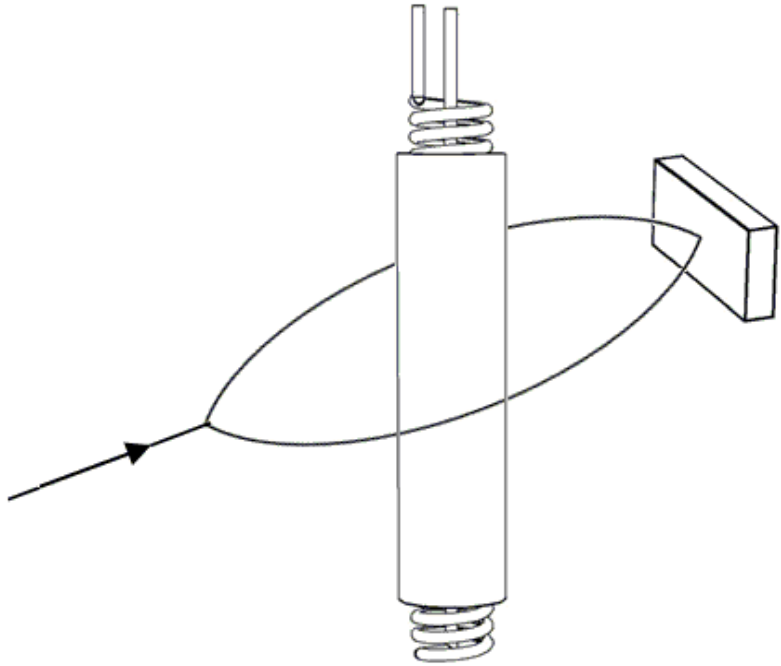

Fig. 2, A A-B effect experiment with the solenoid coated by a superconducting cylinder.

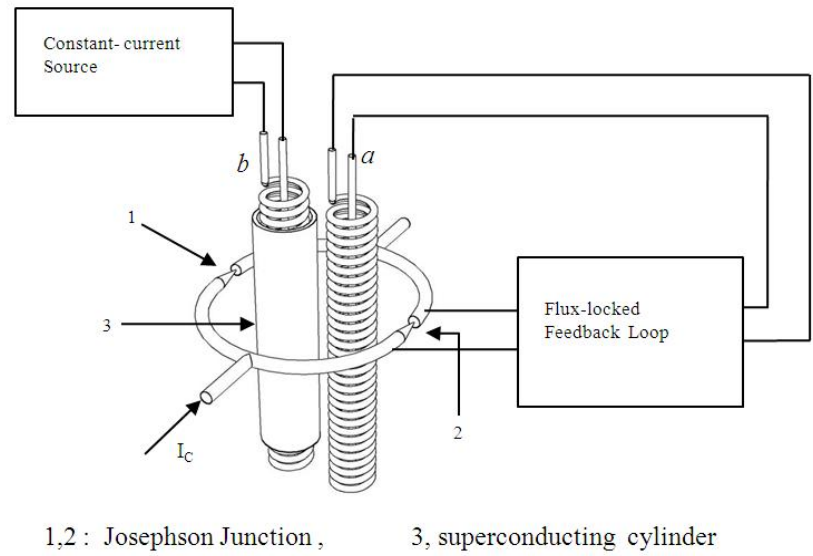

Fig. 3, A new experimental scheme to test the interpretations on the A-B effect.